\documentclass[article,aps,nofootinbib,showpacs,amsfonts,epsf]{revtex4}

\input{epsf.tex}

\newcommand{\E}{{\cal{E}}}

\newcommand{\s}{\sigma}

\renewcommand{\d}{{\rm d}}

\newcommand{\be}{\begin{equation}}
\newcommand{\ee}{\end{equation}}
\newcommand{\bea}{\begin{eqnarray}}
\newcommand{\eea}{\end{eqnarray}}
\newcommand{\ba}{\begin{array}}
\newcommand{\ea}{\end{array}}
\def\J#1#2#3#4{#1 {\it #2} {\bf #3} #4}
\def\PRD{Phys. Rev. D}
\def\PR{Phys. Rev.}
\def\PRL{Phys. Rev. Lett.}
\def\PTP{Prog. Theor. Phys.}
\def\APN{Ann. Phys. (NY)}

\def\AP{Ann. Physik}

\def\JMP{J. Math. Phys.}

\def\JHEP{J. High Energy Phys.}

\def\CQG{Class. Quantum Grav.}

\def\PLA{Phys. Lett. A}

\def\NPB{Nucl. Phys. B}

\def\NC{Nuovo Chim.}

\begin{document}
\draft

\title{The Bret\'on-Manko equatorially antisymmetric binary configuration revisited}

\author{V.~S.~Manko$^\dag$, R. I. Rabad\'an$^\dag$ and E.~Ruiz$\,^\ddag$}
\address{$^\dag$Departamento de F\'\i sica, Centro de Investigaci\'on y de
Estudios Avanzados del IPN, A.P. 14-740, 07000 M\'exico D.F.,
Mexico\\ $^\ddag$Instituto Universitario de F\'{i}sica Fundamental y Matem\'aticas, Universidad de Salamanca, 37008 Salamanca, Spain}

\begin{abstract}
The Bret\'on-Manko solution for two identical counter-rotating
Kerr-Newman charged masses is rewritten in the physical
parametrization involving Komar quantities. The new form of the
solution turns out to be very convenient for verifying that the
black-hole sector of the Bret\'on-Manko binary configuration
saturates a recent geometric inequality for interacting black holes
with struts discovered by Gabach Clement.  \end{abstract}

\pacs{04.20.Jb, 04.70.Bw, 97.60.Lf}

\maketitle


\section{Introduction}

The Bret\'on-Manko (BM) stationary axisymmetric electrovac solution
of the Einstein-Maxwell equations \cite{BMa} was constructed in 1995
with the aid of Sibgatullin's integral method \cite{Sib,MSi}. It is
asymptotically flat and describes a system of two identical
counter-rotating Kerr-Newman (KN) sources, black holes or
hyperextreme objects. The sources in the BM configuration are
prevented from falling onto each other by a massless strut
\cite{Isr}, the latter representing a specific line source of
pressure, which is removable only in the special case of equal
masses and charges when the solution becomes a specialization of the
Parker-Ruffini-Wilkins hyperexteme field \cite{PRW} generated by
means of the Perj\'es-Israel-Wilson method \cite{Per,IWi}. The BM
solution gave rise to a systematic study of equatorially
antisymmetric spacetimes \cite{EMR1,EMR2,SRo} which were shown to
constitute a rather broad and interesting subfamily of stationary
axisymmetric fields. At the same time, the BM solution has also
enlarged a class of metrics able to describe the interacting black
holes -- up to date the particular members of that class have been
obtained and analyzed mainly in the context of the static vacuum
\cite{IKh,CPe,ERe}, electrostatic \cite{CPe,ETe,Man,CMR,CDD} or
stationary vacuum \cite{MRRS,HRe,HRR} spacetimes with struts. While
the paper \cite{BMa} basically deals with the mathematical structure
of the BM metric and associated equilibrium problem, it would
certainly be of interest to better explore the physical properties
of the spacetime whose subextreme sector provides excellent
opportunities for the study of two interacting KN black holes and
for extending the recent results on interacting Kerr sources
\cite{MRRS} to the case of non-vanishing electromagnetic field. It
would be also desirable to use the general BM metric for testing the
newly proposed geometric inequality \cite{Gab} for multiple black
holes with struts.

To make the BM solution more suitable for the physical analysis, in
the present paper we shall rewrite it in a new parametrization
involving the physical Komar quantities \cite{Kom} and a separation
distance as arbitrary parameters. This will enable us to demonstrate
analytically that each KN black-hole constituent from the subextreme
sector of the BM solution verifies the well-known Smarr's mass
formula \cite{Sma}. We will subsequently apply the BM solution for
demonstrating that it saturates the inequality on the bounds of the
interaction force in black-hole systems with struts recently derived
by Gabach Clement \cite{Gab}.

\section{The BM solution in physical parameters}

The original BM solution arises from the axis data
\bea \E(\rho=0,z)=\frac{(z-k-M-ia')(z+k-M+ia')} {(z-k+M-ia')(z+k+M+ia')}, \nonumber\\
\Phi(\rho=0,z)=\frac{2Qz} {(z-k+M-ia')(z+k+M+ia')}, \label{axis_data} \eea
representing the axis expressions of the Ernst complex potentials
\cite{Ern}. The parameters $M$, $a'$, $Q$ and $k$ in
(\ref{axis_data}) are related, respectively, to the mass, angular
momentum, charge of each KN source and to the separation distance,
$\rho$ and $z$ are the Weyl-Papapetrou cylindrical coordinates. In
the paper \cite{BMa}, Sibgatullin's method was applied to the data
(\ref{axis_data}) for obtaining the corresponding Ernst potentials
in the whole space, as well as the metric functions $f$, $\gamma$
and $\omega$ entering the line element
\be
d s^2=f^{-1}[e^{2\gamma}(d\rho^2+d z^2)+\rho^2 d\varphi^2]-f(d
t-\omega d\varphi)^2. \label{Papa} \ee
The explicit original form of $\E$, $\Phi$, $f$, $\gamma$ and
$\omega$ of the BM solution the reader may find in \cite{BMa}.

It should be pointed out that the parametrization used in \cite{BMa}
is not the most attractive one from the physical point of view
because only the parameters $M$ and $Q$ coincide exactly with the
physical mass and charge of each KN constituent; at the same time,
the parameter $a'$ does not coincide with the angular momentum per
unit mass of any of the constituents, whereas the parameter $k$ is
not exactly half the coordinate distance between the centers of the
constituents. In the absence of the electric charge $Q$ (pure vacuum
case) when the BM solution represents two counter-rotating Kerr
sources \cite{Ker} a similar problem was already solved in the paper
\cite{MRRS} by introducing the Komar angular momentum per unit mass
$a$ instead of $a'$, and the coordinate distance $R$ between the
centers of the sources instead of $k$ (see Fig.~1).

\begin{figure}[htb]
\centerline{\epsfysize=90mm\epsffile{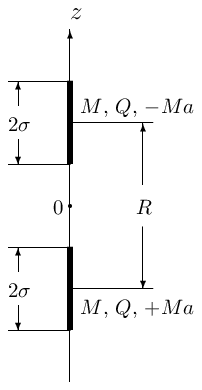}} \caption{Location of
identical counter-rotating KN black holes on the symmetry axis and
the new parametrization.}
\end{figure}

Remarkably, transition from the parameters $(a',k)$ to the
physically more transparent parameters $(a,R)$ can be performed in
the general BM solution too, and after a considerable effort we have
eventually been able to find the required reparametrized form of the
data (\ref{axis_data}) and the form of an important quantity $\s$
generalizing the analogous quantity in the vacuum case (cf. formula
(3) of \cite{MRRS}):
\bea \E(\rho=0,z)=\frac{z^2-2Mz-({\textstyle\frac12}MR+M^2-
{\textstyle\frac12}Q^2-iMa)^2\mu} {z^2+2Mz-
({\textstyle\frac12}MR+M^2- {\textstyle\frac12}Q^2-iMa)^2\mu},
\nonumber\\
\Phi(\rho=0,z)=\frac{2Qz}
{z^2+2Mz-({\textstyle\frac12}MR+M^2-{\textstyle\frac12}Q^2-iMa)^2\mu},
\label{axis_phys} \eea
and
\be \sigma=\sqrt{M^2-Q^2-M^2a^2\mu}, \quad
\mu:=\frac{R^2-4M^2+4Q^2}{(MR+2M^2-Q^2)^2}. \label{sigma} \ee
The new parameters $R$ and $a$ are related to $a'$, $k$, $M$ and $Q$
by the formulae
\bea R&=&2\kappa_+, \quad
a=\frac{a'k(2M^2-Q^2+2M\kappa_+)}{M(M^2-k^2+a'^2+4Q^2-d)}, \nonumber\\
\kappa_+&=&\sqrt{(k^2+M^2-a'^2-2Q^2+d)/2}, \quad
d=\sqrt{(k^2-M^2+a'^2)^2+4M^2a'^2}. \label{Ra} \eea

The application of Sibgatullin's method to the data
(\ref{axis_phys}) yields the following new expressions for the Ernst
potentials of the BM solution instead of the old ones:
\bea \E&=&\frac{A-B}{A+B}, \quad \Phi=\frac{C}{A+B}, \nonumber\\
A&=&(M^2-Q^2)[4\sigma^2(R_+R_-+r_+r_-)+R^2(R_+r_++R_-r_-)]
+[\sigma^2(R^2-4M^2+4Q^2)  \nonumber\\
&&-M^2a^2R^2\mu](R_+r_-+R_-r_+)
-2iaMR\mu\sigma(MR+2M^2-Q^2)(R_+r_--R_-r_+),
\nonumber\\
B&=&2MR\sigma\{R\sigma
(R_++R_-+r_++r_-)-\beta(R_+-R_--r_++r_-)\},
\nonumber\\
C&=&QB/M, \nonumber\\
R_\pm&=&\sqrt{\rho^2+(z+{\textstyle\frac{1}{2}}R\pm\sigma)^2},
\quad r_\pm=\sqrt{\rho^2+(z-{\textstyle\frac{1}{2}}R\pm\sigma)^2}, \quad \beta:=2(M^2-Q^2)+iMa\mu(MR+2M^2-Q^2).
\label{EP}  \eea
On the other hand, the new expressions for the functions $f$,
$\gamma$ and $\omega$ of the BM solution have been found to have the
form
\bea f&=&\frac{AA^*-BB^*+CC^*}{(A+B)(A^*+B^*)}, \,\,\,
e^{2\gamma}=\frac{AA^*-BB^*+CC^*}{16R^4\sigma^4R_+R_-r_+r_-}, \,\,\,
\omega=-\frac{{\rm Im}[2G(A^*+B^*)+CI^*]}{AA^*-BB^*+CC^*},
\nonumber\\
G&=&-zB+R\sigma\{(2M^2-Q^2)[2\sigma(r_+r_--R_+R_-)+R(R_-r_--R_+r_+)] \nonumber\\ 
&&+M(R+2\sigma)(R\sigma-\beta)(R_+-r_-)+M(R-2\sigma)(R\sigma+\beta)(R_--r_+)\},
\nonumber\\
I&=&\frac{Q}{M}\{G+RQ^2\sigma[2\sigma(r_+r_--R_+R_-)+R(R_-r_--R_+r_+)]\},
\label{metric} \eea
where an asterisk denotes complex conjugation and ${\rm Im}(x)$ the
imaginary part of $x$. Notice that the metric function $\omega$ in
(\ref{metric}) is presented in a simpler and more elegant form than
in the paper \cite{BMa}.

For completeness, below we also write out expressions for the
potentials $A_4$ and $A_3$ of the BM solution which represent,
respectively, the electric and magnetic components of the
electromagnetic 4-potential $A_i=(0,0,A_3,A_4)$:
\be A_4={\rm Re}\left(\frac{C}{A+B}\right), \quad A_3={\rm
Im}\left(\frac{I}{A+B}\right), \label{em_pot} \ee
where ${\rm Re}(x)$ means taking real part of $x$. The knowledge of
the above $A_4$ and $A_3$ is needed in particular for the analysis
of Smarr's mass formula.

\section{Some physical properties of BM solution}

The real valued $\s$ determine a black-hole sector of the BM
solution. In this case there are two horizons\footnote{They are in
fact Killing horizons by the same argument as used in \cite{DHo} for
the case of the double-Kerr solution \cite{KNe} (see p.~348 of
\cite{DHo}).} determined by the null hypersurfaces $\rho=0$,
${\textstyle\frac12}R-\s\le z\le {\textstyle\frac12}R+\s$ and
$\rho=0$, $-{\textstyle\frac12}R-\s\le z\le
-{\textstyle\frac12}R+\s$, which in the cylindrical coordinates
($\rho,z$) are represented by rods of length $2\s$ (see Fig.~1).
Each KN constituent then must verify Smarr's mass formula for black
holes
\be M=\frac{1}{4\pi}\kappa S+2\Omega^H J+\Phi^H Q, \label{Sma} \ee
which, apart from the mass $M$, angular momentum $J$ and charge $Q$,
contains four quantities to be evaluated on the horizon: the surface
gravity $\kappa$, the area of the horizon $S$, horizon's angular
velocity $\Omega^H$ and the electric scalar $\Phi^H$. Since the
constituents in the BM solution only differ in the orientation of
their angular momenta, it is sufficient to check formula (\ref{Sma})
for solely one subextreme constituent, say, the upper one; the
second constituent will then have opposite angular momentum $-J$ and
horizon's velocity $-\Omega^H$ whose product leaves invariant the
second term on the right hand side of (\ref{Sma}). The known Komar
characteristics associated with the upper constituent are $M$,
$J=-Ma$, $Q$, and this fact can be checked straightforwardly by
using Tomimatsu's formulae \cite{Tom}
\bea M&=&-\frac{1}{8\pi}\int_{H}
\omega\Omega_{,z}\d\varphi\d z, \nonumber\\
J&=&\frac{1}{8\pi}\int_{H}\omega
\left[-1-{\textstyle\frac12}\omega\Omega_{,z}+\tilde
A_3A'_{3,z}+(A_3A'_3)_{,z}\right]\d\varphi\d z, \nonumber\\
Q&=&\frac{1}{4\pi}\int_{H} \omega A'_{3,z}\d\varphi\d z, \label{kq}
\eea
where $\Omega={\rm Im}(\E)$, $A'_3={\rm Im}(\Phi)$, $\tilde
A_3=A_3-\omega A_4$, and the integration must be performed over the
horizon $H$ on which the metric function $\omega$ and the potential
$\tilde A_3$ take constant values. Obviously, the total mass, total
angular momentum and total charge of the BM system are equal,
respectively, to $2M$, $0$ and $2Q$.

On the other hand, the quantities $\kappa$, $S$, $\Omega^H$ and
$\Phi^H$ can be calculated with the aid of the formulae
\cite{Car,Tom}
\be \kappa=\sqrt{-\omega^{-2}e^{-2\gamma}}, \quad
S=\frac{4\pi\sigma}{\kappa}, \quad \Omega^H=\omega^{-1}, \quad
\Phi^H=A_4-\Omega^H A_3, \label{kap} \ee
thus yielding in our case
\bea \kappa&=&\frac{R\s}{\Delta}, \quad S=\frac{4\pi\Delta}{R},
\quad \Omega^H=-\frac{Ma\mu(MR+2M^2-Q^2)}{\Delta}, \nonumber\\
\Phi^H&=&\frac{Q[(R+2M)(M+\s)-2Q^2]}{\Delta}, \nonumber\\
\Delta&=&2M(R+2M)(M+\s)-Q^2(R+4M+2\s). \label{kap_res} \eea
A direct check shows that the above quantities satisfy identically
the relation (\ref{Sma}), therefore the subextreme constituents in
the BM solution are indeed charged rotating black holes.

Mention that in the limit $R\to\infty$ (infinite separation) one
recovers from (\ref{kap_res}) the characteristics of a single KN
black hole \cite{NCC}, and at finite $R$, formulae (\ref{kap_res})
take into account the interaction between the black holes. The
interaction force in the BM configuration is determined by a very
simple expression ($\gamma_0$ is the value of the metric function
$\gamma$ on the strut)
\be {\cal F}=\frac14(e^{-\gamma_0}-1)=\frac{M^2-Q^2}{R^2-4M^2+4Q^2},
\label{F} \ee
which coincides with the expression for the interaction force
between two identical Reissner-Nordstr\"om non-rotating charged
masses \cite{Rei,Nor} (cf. formula (30) of \cite{Man}). Mention that
geometrically ${\cal F}$ can be defined in terms of the conical
deficit $\delta$ associated with the strut as ${\cal
F}=-\delta/(8\pi)$. From (\ref{F}) and (\ref{sigma}) one easily
infers that no equilibrium states of two identical counter-rotating
black holes with non-zero angular momenta are possible without a
supporting strut; at the same time, when $M^2=Q^2$, we have either
equilibrium of two hyperextreme KN constituents ($a\ne 0$), or that
of two Reissner-Nordstr\"om extreme black holes ($a=0$).

The limiting case of two extreme counter-rotating black holes is
defined by the condition $\s=0$. Solving the latter for $R$, we find
that $M$ and $Q$ must satisfy the inequality $M^2>Q^2$ as a
condition of the reality of $R$ and non-vanishing $a$. On the other
hand, solving $\s=0$ for $a$, we get
\be a^2=\frac{M^2-Q^2}{M^2\mu}, \label{ec_a} \ee
and taking into account that $0<M^2\mu<1$ for all $M>0$,
$R>2\sqrt{M^2-Q^2}$, we conclude that a black hole in the BM
configuration needs a larger absolute value of the angular momentum
in order to become extremal than a single KN black hole. The Ernst
potentials and metric functions of the extreme BM solution have been
worked out in explicit form in the paper \cite{MRS}.

Let us also comment, in connection with recent interest in
geometrical inequalities for black holes (we refer the reader to a
comprehensive review \cite{Dai} on the topic), that the extreme BM
solution provides the first example of a stationary axisymmetric
electrovac spacetime for which the equality is achieved in a recent
relation for black holes with struts derived by Gabach Clement
\cite{Gab}:
\be \sqrt{1+4{\cal F}}\ge\frac{\sqrt{(8\pi J)^2+(4\pi Q^2)^2}}{S}. \label{Gab} \ee
Indeed, taking into account (\ref{ec_a}), the extremal values of $S$
and $(8\pi J)^2+(4\pi Q^2)^2$ are
\bea  &&S=\frac{4\pi}{R}[(M^2-Q^2)(R+4M)+M^2R], \nonumber\\
&&(8\pi J)^2+(4\pi Q^2)^2
=\frac{16\pi^2[(M^2-Q^2)(R+4M)+M^2R]^2}{R^2-4M^2+4Q^2},
\label{S_ext} \eea
with which, and with formula (\ref{F}), the equality in (\ref{Gab})
can be easily verified. In general, the inequality (\ref{Gab})
establishes lower bounds on the interaction force between the
black-hole constituents in the BM solution.

We conclude this section by noticing that Smarr's formula gives rise
to the inequality
\be M\ge 2\Omega^H J+\Phi^H Q \label{Sma_in} \ee
since the first term on the right hand side of (\ref{Sma}) is a
non-negative quantity. Then the equality in (\ref{Sma_in}) will be
achieved exclusively by the extremal black holes, while the
sub-extreme black holes will verify a strict inequality.

\section{Conclusions}

The new representation of BM solution elaborated in the present
paper is remarkable in several aspects: it is the first and most
simple example of a stationary axisymmetric electrovac spacetime
entirely written in physical parameters and able to describe the
field of two interacting KN black holes (previous results on the
physical parametrization of binary configurations were obtained
either for a vacuum \cite{MRRS} or static electrovac \cite{Man}
cases). It considerably simplifies the analysis of physical
properties of the BM solution and permits one to get concise
expressions for the basic characteristics of interacting KN black
holes and use them in concrete applications, in particular, for
proving that the BM binary configuration saturates the
Gabach-Clement inequality for interacting black holes with struts --
one of the main results of the present paper. The new form of the BM
metric also motivates the search for physical parametrizations of
more general configurations of KN black holes than the BM one, for
instance by introducing appropriately the charges into the vacuum
solution for non-equal counter-rotating Kerr black holes \cite{CLM}.
Last but not least, an interesting aspect of the reparametrized BM
solution is connected with a recent numerical study by Dain and
Ortiz \cite{DOr} of the non-stationary axisymmetric binary
black-hole systems using a procedure that employs stationary
axisymmetric spacetimes with struts as ``momentary stationary
data'', thus suggesting that the BM solution could in principle be
used in the modern numerical codes for obtaining new information
about the non-radiating binary non-stationary systems too.

\section*{Acknowledgements}
We are grateful to the anonymous referee for many valuable comments
and suggestions that helped us to considerably improve the
presentation. We also thank Erasmo G\'omez for technical computer
support during the involved analytic calculations. This work was
partially supported by CONACYT, Mexico, and by Ministerio de Ciencia
y Tecnolog\'\i a, Spain, under the Project FIS2009-07238.


\begin{references}

\bibitem{BMa} Bret\'on N and Manko V S \J{1995}{\CQG}{12}{1969}

\bibitem{Sib} Sibgatullin N R 1991 \emph{Oscillations and Waves in Strong
Gravitational and Electromagnetic Fields} (Berlin: Springer)

\bibitem{MSi} Manko V S and Sibgatullin N R \J{1993}{\CQG}{10}{1383}

\bibitem{Isr} Israel W \J{1977}{\PRD}{15}{935}

\bibitem{PRW} Parker L, Ruffini R and Wilkins D \J{1973}{\PRD}{7}{2874}

\bibitem{Per} Perj\'es Z \J{1971}{\PRL}{27}{1668}

\bibitem{IWi} Israel W and Wilson G A \J{1972}{\JMP}{13}{865}

\bibitem{EMR1} Ernst F J, Manko V S and Ruiz E \J{2006}{\CQG}{23}{4945}

\bibitem{EMR2} Ernst F J, Manko V S and Ruiz E \J{2007}{\CQG}{24}{2193}

\bibitem{SRo} Sod-Hoffs J and Rodchenko E D \J{2007}{\CQG}{24}{4617}

\bibitem{IKh} Israel W and Khan K A \J{1964}{\NC}{33}{331}

\bibitem{CPe} Costa M S and Perry M J \J{2000}{\NPB}{591}{469}

\bibitem{ERe} Emparan R and Reall H S \J{2002}{\PRD}{65}{084025}

\bibitem{ETe} Emparan R and Teo E \J{2001}{\NPB}{610}{190}

\bibitem{Man} Manko V S \J{2007}{\PRD}{76}{124032}

\bibitem{CMR} Chng B, Mann R, Radu E and Stelea C \J{2008}{\JHEP}{12}{009}

\bibitem{CDD} Stelea C, Dariescu C and Dariescu M \J{2011}{\PRD}{84}{044009}

\bibitem{MRRS} Manko V S, Rodchenko E D, Ruiz E and Sadovnikov B I \J{2008}{\PRD}{78}{124014}

\bibitem{HRe} Herdeiro C A R and Rebelo C \J{2008}{\JHEP}{10}{017}

\bibitem{HRR} Herdeiro C A R, Radu E and Rebelo C \J{2010}{\PRD}{81}{104031}

\bibitem{Gab} Gabach Clement M E \J{2012}{\CQG}{29}{165008}

\bibitem{Kom} Komar A \J{1959}{\PR}{113}{934}

\bibitem{Sma} Smarr L \J{1973}{\PRL}{30}{71}

\bibitem{Ern} Ernst F J \J{1968}{\PR}{168}{1415}

\bibitem{Ker} Kerr R P \J{1963}{\PRL}{11}{237}

\bibitem{DHo} Dietz W and Hoenselaers C \J{1985}{\APN}{165}{319}

\bibitem{KNe} Kramer D and Neugebauer G \J{1980}{\PLA}{75}{259}

\bibitem{Car} Carter B 1979 in {\it General Relativity, an Einstein Centenary Survey},
ed. S. W. Hawking and W. Israel (Cambridge University Press,
Cambridge), p. 294.

\bibitem{Tom} Tomimatsu A \J{1984}{\PTP}{72}{73}

\bibitem{NCC} Newman E T, Couch E, Chinnapared K, Exton A, Prakash A and Torrence R
\J{1965}{\JMP}{6}{918}

\bibitem{Rei} Reissner H \J{1916}{\AP}{50}{106}

\bibitem{Nor} Nordstr\"om G 1918 {\it Proc. Kon. Ned. Akad. Wet.} {\bf 20} 1238

\bibitem{MRS} Manko V S, Ruiz E and Sadovnikova M B \J{2011}{\PRD}{84}{064005}

\bibitem{Dai} Dain S \J{2012}{\CQG}{29}{073001}

\bibitem{CLM} Cabrera-Munguia I, L\"ammerzahl C and Mac\'ias A 2013 {\it Preprint} gr-qc/1302.4843

\bibitem{DOr} Dain S and Ortiz O E \J{2009}{\PRD}{80}{024045}

\end{references}
\end{document}